# COVID-19 Antibody Test / Vaccination Certification
## There's an app for that

Marc Eisenstadt, Manoharan Ramachandran, Niaz Chowdhury, Allan Third, John Domingue*

## Visual Summary / Graphical Abstract + Legend
*(Rotated/enlarged image overleaf, followed by full paper and Supplementary Materials)*

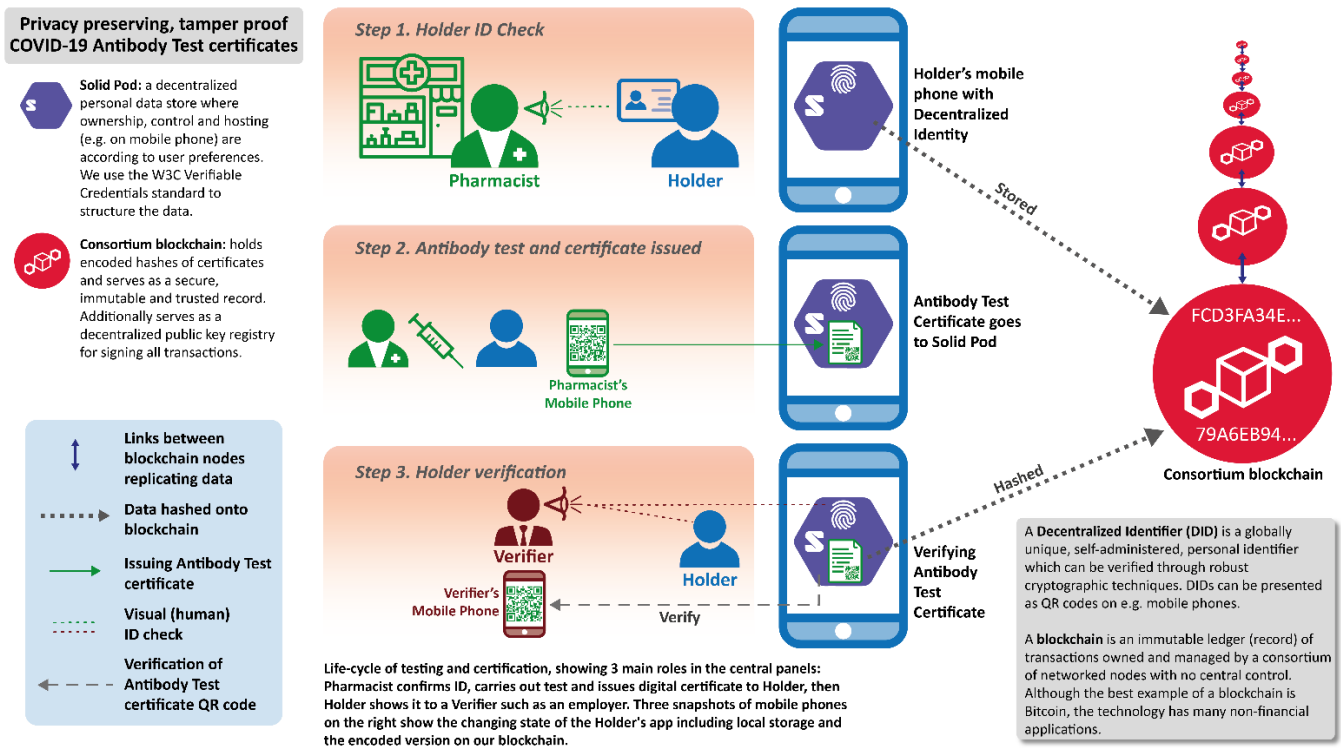

This work addresses the issues involved in providing robust certification for COVID-19 immunity (assuming the biological premise of 'immunity' is ultimately confirmed). Methods: We developed a prototype mobile phone app and scalable distributed server architecture that facilitates instant verification of tamper-proof test results. Personally identifiable information is only stored at the user's discretion, and the app allows the end-user selectively to present only the specific test result with no other personal information revealed. Behind the scenes it relies upon (a) the 2019 World Wide Web Consortium standard called 'Verifiable Credentials', (b) Tim Berners-Lee's decentralized personal data platform 'Solid', and (c) a consortium Ethereum-based blockchain. Results: Our architecture enables verifiability and privacy in a manner derived from public/private key pairs and digital signatures, generalized to avoid restrictive ownership of sensitive digital keys and/or data. Benchmark performance tests show it to scale linearly in the worst case, as significant processing is done locally on each app. For the test certificate Holder, Issuer (e.g. doctor, pharmacy) and Verifier (e.g. employer), it is 'just another app' which takes only minutes to use. Conclusions: The app and distributed server architecture offer a prototype proof of concept that is readily scalable, widely applicable to personal health records and beyond, and in effect 'waiting in the wings' for the biological issues, plus key ethical issues raised in the discussion section, to be resolved.

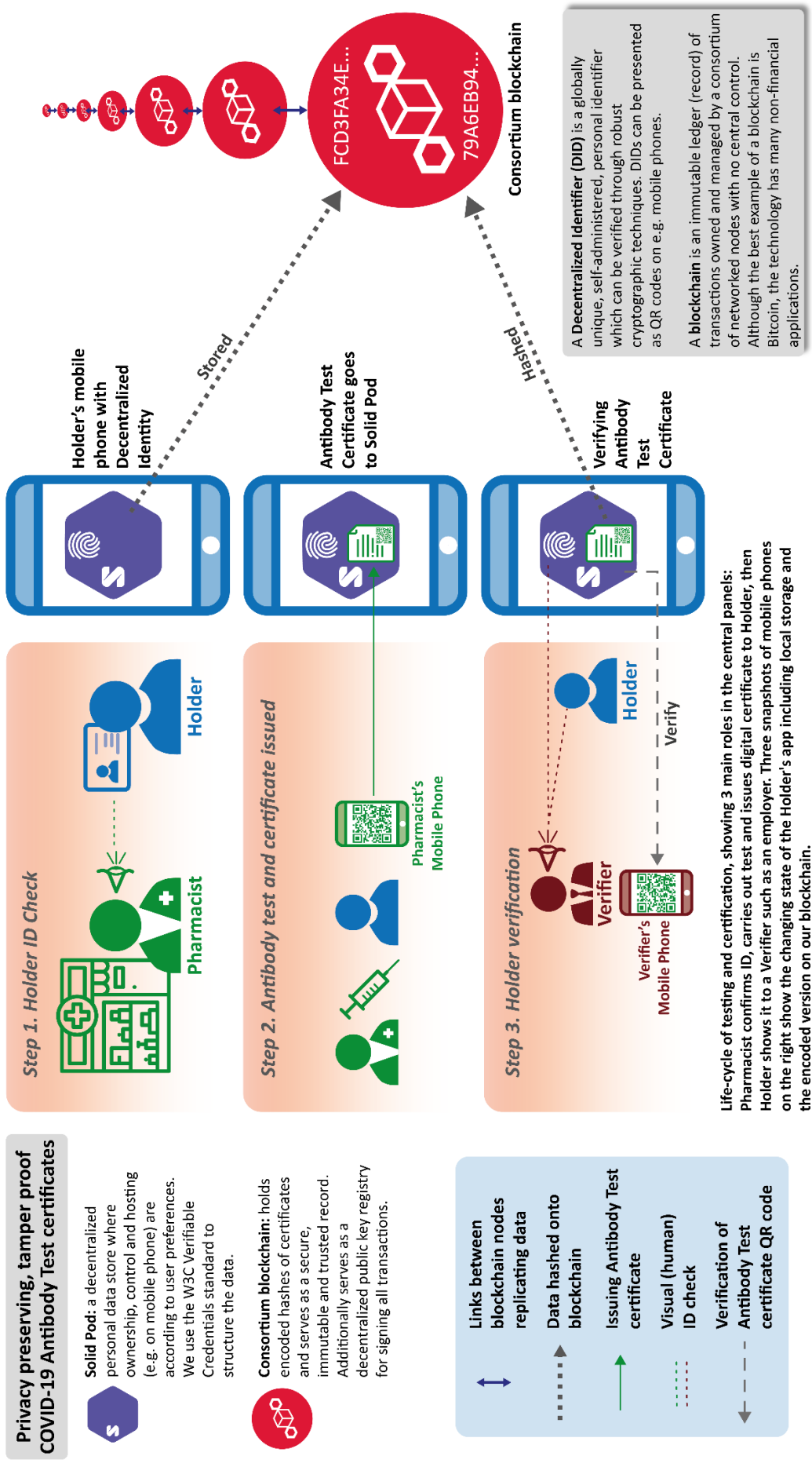

# COVID-19 Antibody Test / Vaccination Certification: There's an app for that

Marc Eisenstadt, Manoharan Ramachandran, Niaz Chowdhury, Allan Third, John Domingue*

**Full Paper Citation:** M. Eisenstadt, M. Ramachandran, N. Chowdhury, A. Third and J. Domingue, "COVID-19 Antibody Test/Vaccination Certification: There's an App for That," in *IEEE Open Journal of Engineering in Medicine and Biology*, vol. 1, pp. 148-155, 2020, doi: 10.1109/OJEMB.2020.2999214.

# COVID-19 Antibody Test/Vaccination Certification: There's an App for That

Marc Eisenstadt, Manoharan Ramachandran, Niaz Chowdhury, Allan Third, and John Domingue

*Abstract—Goal:* As the Coronavirus Pandemic of 2019/2020 unfolds, a COVID-19 'Immunity Passport' has been mooted as a way to enable individuals to return back to work. While the quality of antibody testing, the availability of vaccines, and the likelihood of even attaining COVID-19 immunity continue to be researched, we address the issues involved in providing tamper-proof and privacy-preserving certification for test results and vaccinations. *Methods:* We developed a prototype mobile phone app and requisite decentralized server architecture that facilitates instant verification of tamper-proof test results. Personally identifiable information is only stored at the user's discretion, and the app allows the end-user selectively to present only the specific test result with no other personal information revealed. The architecture, designed for scalability, relies upon (a) the 2019 World Wide Web Consortium standard called 'Verifiable Credentials', (b) Tim Berners-Lee's decentralized personal data platform 'Solid', and (c) a Consortium Ethereum-based blockchain. *Result*s: Our mobile phone app and decentralized server architecture enable the mixture of verifiability and privacy in a manner derived from public/private key pairs and digital signatures, generalized to avoid restrictive ownership of sensitive digital keys and/or data. Benchmark performance tests show it to scale linearly in the worst case, as significant processing is done locally on each app. For the test certificate Holder, Issuer (e.g. healthcare staff, pharmacy) and Verifier (e.g. employer), it is 'just another app' which takes only minutes to use. *Conclusions:* The app and decentralized server architecture offer a prototype proof of concept that is readily scalable, applicable generically, and in effect 'waiting in the wings' for the biological issues, plus key ethical issues raised in the discussion section, to be resolved.

*Impact Statement*—As soon as COVID-19 antibody testing, vaccines, and likelihood of immunity surpass quality thresholds, our tamper-proof and privacy-preserving certification can be rapidly deployed. Our approach is applicable to any certification scenario.

## I. INTRODUCTION

THE Coronavirus/COVID-19 pandemic of 2019/2020 is still taking its terrible toll as we write this [1]. Tests for the presence of antibodies *could* offer a way for people who can prove COVID-19 immunity to go back to work [2], [3]. There are, however, challenges concerning the biological premise of 'immunity': the strength and longevity of COVID-19 immunity after infection are matters of current debate and research, as are the sensitivity and robustness of the relevant tests [4], [5] and the race to develop a viable vaccine [6], [7].

Given the scale of the pandemic and financial fallout, it is plausible that 'COVID-19 antibody test / vaccination certification' (henceforth 'CAT/VC'), if shown to be robust, will be in great demand. Bearing in mind the legal and ethical implications of such certification, raised in [8], [9] and our Discussion, we feel that for either the current pandemic or a pandemic of the future, the concept of certification has a place, *particularly when the recipient is employed in healthcare or other key sectors*.

But what form should certification take? A signed or stamped letter is the centuries-old default, and straightforward to roll out at scale, as long as there is some point-of-test proof of identity. Our approach is based on the view that for such a sensitive and likely high-value certificate, a paper version is too vulnerable to alteration or forgery (an exception arises in environments that are 'lower tech' for socio-economic reasons and we later describe a printed certificate to address this case). A digital certificate makes the most sense, provided that it can be: (i) Privacy-preserving (because as proud as the holder might be of new-found 'immunity', personal data can be re-purposed in unpredictable ways [10]), (ii) un-forgeable, (iii) easy to administer, (iv) easily verifiable while still preserving privacy, (v) scalable to millions of users, and (vi) cost-effective.

All of this effort would be wasted without public acceptance, which is increasingly challenging in an era of suspicion about data-collecting apps [11]. Toward this end, we argue not only for the decentralized approach underlying our design and implementation below, but also for its benefits in allowing individuals who have been tested to change their minds and quit the scheme, knowing that even cryptographically encoded data will be 'orphaned' (no data pointing to it), rendering it meaningless. Also,

Manuscript received April 20, 2020; revised May 22, 2020 and May 28, 2020; accepted May 28, 2020. Date of publication June 1, 2020; date of current version June 26, 2020. This work was supported by the UK Government Office for Students' Institute of Coding and two projects funded under the European Union's Horizon 2020 research and innovation programme: QualiChain (grant agreement number 822404) and DEL4ALL (871573). *(Corresponding author: John Domingue.)*
The authors are with the Knowledge Media Institute, The Open University, Milton, Keynes MK7 6AA, U.K. (e-mail: marc.eisenstadt@open.ac.uk; manoharan.ramachandran@open.ac.uk; niaz.chowdhury@open.ac.uk; allan.third@open.ac.uk; john.domingue@open.ac.uk).
This article has supplementary downloadable material available at https://ieeexplore.ieee.org, provided by the authors.
Digital Object Identifier 10.1109/OJEMB.2020.2999214

in the Supplementary Materials, we emphasize the importance of having strong oversight by an ethics watchdog to ensure best endeavours to avoid unleashing a Pandora's Box of undesirable side-effects.

How best to undertake such a challenge? Modern smartphone apps and several key technologies such as public key cryptosystems and immutable blockchain records offer some tantalizing prospects for the path we envisage, if they can satisfy the above criteria. Below, we look at the methods by which this can be achieved, assuming a scenario involving testing by a known authority (e.g. a healthcare practitioner or pharmacist), as opposed to self-testing at home. This main paper assumes an 'On-Site Test for Antibodies + Issuance of Digital Certificate Including Photo ID' in order to explain our approach, and in the Supplementary Materials we describe variations for (a) 'Issuing Digital Certificate Without Photo ID', (b) 'Issuing Paper Certificate', (c) 'Off-Site Testing Via External Lab', and (d) 'Vaccination + Certification'.

## II. METHODS

We focus on the design and implementation of a prototype mobile phone app and requisite decentralized server architecture, intended to facilitate verification of tamper-proof test results. Our design involves a novel hybrid architecture based on (a) the 2019 World Wide Web Consortium standard called 'Verifiable Credentials', (b) Tim Berners-Lee's decentralized personal data platform 'Solid', and (c) a Consortium Ethereum-based blockchain. We work through (d) a plausible use case scenario, then (e) describe the key 'onboarding' and certification steps in detail; and (f) provide benchmark tests to anticipate scaling performance.

### A. 'Verifiable Credentials' For Digital Certification

Verifiable Credentials [12] is a W3C standard that builds upon Public Key Infrastructure (PKI), the public/private key pairs that facilitate digital signatures in widespread use today. The W3C extensions standardize the definitions of document formats to make them machine-readable and communicable, and to generalize PKI, which tends to be costly and highly centralized. The generalization involves a decentralized registry for cryptographic keys, typically residing in a blockchain — this allows every public key to have its own unique address, known as a Decentralized Identifier (DID). The key roles and transactions, adapted for our specific use case, are illustrated in Fig. 1.

The 'Issuer', in our case a trusted pharmacy or the UK National Health Service (NHS), can issue credentials such as blood test results and vaccination certificates. 'Holders' (typically citizen end-users) can store them in their own preferred way, for example in digital wallets that are part of a mobile phone app. 'Verifiers', such as employers, or establishments seeking proof of some attribute, can ask the Holder to present such proof concerning these credentials. Verifiers also check digital signatures against what is known as a 'verifiable (decentralized) data registry': this is the blockchain where the DIDs mentioned above reside.

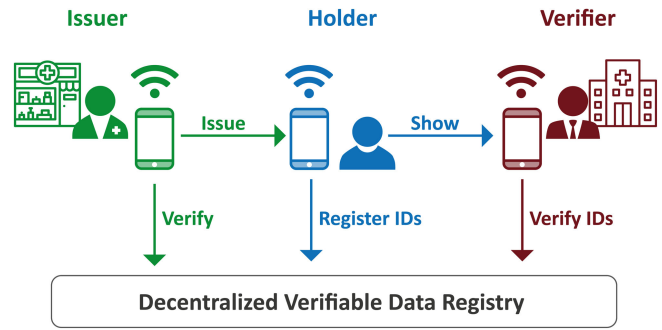

**Fig. 1.** Main roles and workflow in W3C Verifiable Credentials [12], adapted for our COVID-19 Antibody Testing use case.

### B. 'Solid': Decentralized Personal Data

We pointed out in [13] that the over-centralization of data, particularly its consolidation into 'silos' by brand-name IT services and social network providers, is of increasing concern. Decentralization is an ideal starting point for storing sensitive data, including medical, financial, and other personal data — but only if security and privacy are significantly better than what can be offered by traditional centralized systems.

We identified a promising approach to widespread deployment, known as Solid, initiated by Sir Tim Berners-Lee [14], [15]. Solid aims to decentralize the Web by transferring control of data from a central authority to users, thereby allowing users to retain complete ownership of their data, which they store in what are called 'Solid Pods' — analogous to a personal web server that is hosted either locally on a mobile phone, or hosted with a cloud provider of the individual's choice, or both. The key distinction from centralized approaches is that even in the provider-hosted case, the provider's access to the data is limited by the user's preferences.

In [16] we proposed an approach combining Solid Pods and distributed ledgers, of the type familiar to the blockchain community, to facilitate the complete decentralization of data. The key ingredients of this combination are illustrated in Fig. 2, which also provides an overview of the main test/certify/verify life cycle. Our methods give users total control over their data while maintaining the integrity of the stored information through blockchain-based verification.

As in Fig. 1, the 'Holder' is the primary individual who is self-motivated to obtain the certificate of COVID-19 antibody test results in order to be admitted to a workplace or other location. Holders own, manage, and control their own Solid Pods (shown as hexagons in the Holder's mobile phone in Fig. 2 at A, E, and F), which contain their personal data. In Fig. 2, our Holder's Solid Pod contains a elements of a physical ID such as a driving license ('thumbprint' icon at A) and the Holder's signed and countersigned certificate of COVID-19 antibody test results — represented in Fig. 2 as a document in which is embedded a special QR code (F). The Holder is free to store the Solid Pod data on his/her mobile phone, on a personal favorite cloud provider, or both (we only show the mobile phone version for simplicity). At any time, Holders can move or delete data, as it remains under their ownership. One-way encoded 'hashes' of the data (only a few bytes in size) are held, as shown by the dashed arrows in Fig. 2 (E and H), on a blockchain to support independent verification.

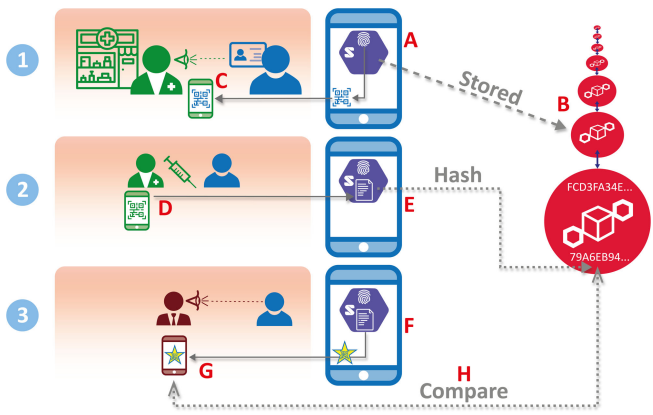

**Fig. 2.** Steps in testing, certification, and verification, showing a Solid Pod (hexagon) hosted on the Holder's mobile phone (labels A, E, F), with minimal hash storage for verification. The circles at B depict replicated blockchain nodes on multiple servers, receding into the distance.

### C. Consortium Blockchain

In our design, we use a 'Consortium blockchain', shown in Fig. 2(B) as circles (depicting multiple replicated blockchain nodes receding into the distance): this is not a fully public blockchain like Ethereum or Bitcoin, but rather a blockchain shared specifically by a *Consortium* of known providers who have signed up to the Ethics Guidelines we describe in the Discussion Section. The Open University-led Consortium blockchain is a private Ethereum network known as OpenEthereum (formerly Parity Ethereum) [17], [18] which uses a 'Proof of Authority' consensus mechanism [19] wherein several nodes can be in the mutually-agreed privileged position of being allowed to confirm transactions. As we go to press, our Consortium blockchain comprises nodes run by The Open University, BT, Condatis, Inrupt, and the Chiba Institute of Technology near Tokyo, with expansion planned as our prototype implementation is scaled up via other large-company partnerships now under discussion. This approach contrasts with that of Bitcoin and other early blockchains which use the slow and ecologically unfriendly Proof of Work, wherein massive computing power enables nodes to have a better chance of confirming transactions. The Consortium approach gives us the kind of distributed scalability that increases security, but without the widespread public availability that may serve as a disincentive for individuals to participate.

### D. Use Case Scenario

In our scenario, the Issuer (Pharmacy) needs to authenticate that the Holder is who they say they are, and thus requests that the Holder display (a) a physical ID, such as a Driving License or a Passport, and (b) a QR code which is scanned by the Issuer using the Issuer's mobile phone app, both of which are shown in Fig. 2 (C). At this point the Issuer taps to accept the ID, and the Holder's photo is 'burned' into the upcoming steps so that at the final step of verification there will be no need to display the same physical ID. The next steps are as follows:

2) The blood test is performed, and the certificate with results is issued as soon as the results are available (off-site lab tests are dealt with in the Supplementary Materials). The Issuer (first scanning a printed QR code if preferred) generates a digitally-signed test result as a new QR code (labelled D in Fig. 2) for transmission to the Holder, thereby providing a Verifiable Credential which is digitally signed by both the Issuer and the Holder, and stored on the Holder's Solid Pod (Fig. 2, D and E). At label E we also see that a hash of the Verifiable Credential is stored on the Consortium blockchain to facilitate verification at step 3.
3) The Holder can now present a provably valid certificate to the Verifier. To avoid someone else impersonating the Holder, the Holder's ID photo was already 'burned' into the digital certificate at Step 1, so the Holder needs to present only the QR code (F and G in Fig. 2)

At H in Fig. 2 we see that the Verifier's app automatically verifies both digital signatures and the certificate against the hash stored on the Consortium blockchain, and confirms acceptance of the COVID-19 Antibody Test Certificate. The certificate stores quantitative test results, such as antibody type (e.g. 'IgG') and level, so it is up to the Verifier's own contextually guided procedures to decide whether to admit the Holder, for example, to work.

### E. Primary design (Onboarding and Certification)

Below we separately describe the details for (i) 'Onboarding' for Issuers, Holders and Verifiers, and (ii) how Certification works behind the scenes. The companion step of (iii) Verification is conceptually similar, and thus provided separately in the Supplementary Materials, as are the more straightforward descriptions of the server and mobile app functional architectures.

*1) Onboarding:* There are three entities involved in the operations: Issuers, Holders and Verifiers. The onboarding process lets all of them install and configure the app. The configuration process for each of them is distinct and requires specific documentation.

**Issuers:** The onboarding of a potential Issuer (Fig. 3) begins with the person downloading and installing the app. The app then instructs the Issuer to complete an in-app form. Because the Issuer has the ability to test, validate and issue certificates to individuals, the app employs *two factor authentication* for all potential Issuers. We anticipate using the API provided by the General Pharmaceutical Council, or an equivalent, to cross-check the registration and the branch information of the likely Issuer (this is simulated in our prototype — discussions about API access are underway), followed by email verification. The former requires the person to input appropriate information into the form, while the latter asks the potential Issuer to provide a valid official email address at the company's registered domain name. The app sends a special link to that Issuer's email address to complete the registration. Data provided by the potential Issuers resides on each Issuer's Solid Pod.

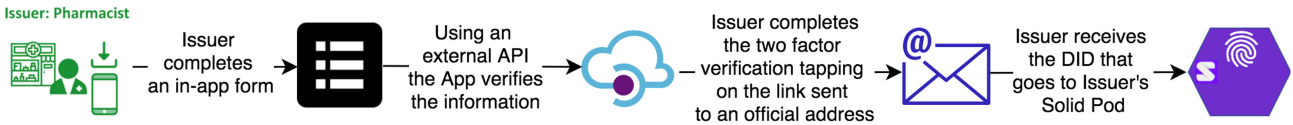

Fig. 3. Issuer onboarding timeline details.

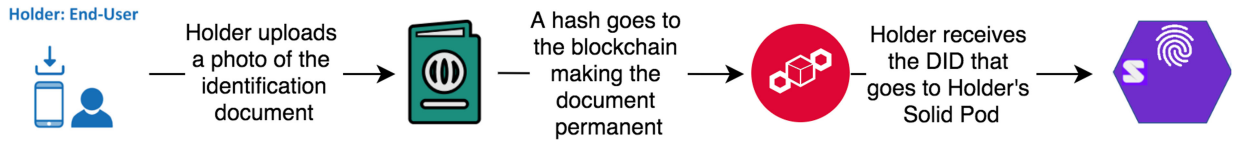

Fig. 4. Holder onboarding timeline details.

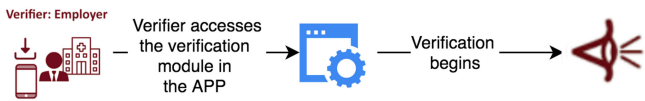

Fig. 5. Verifier onboarding timeline details.

**Holders:** The process of onboarding a Holder (Fig. 4) involves adding an identification document such as a driving license or passport. The document number is used to generate the Decentralized ID (DID) that acts as the anchor for the Holder. A potential Holder first downloads and installs the app followed by adding a photo of the identification document. This document resides in the Holder's Solid Pod. This photo document is deemed permanent (but remains on their personal Solid Pod) and once submitted, cannot be changed again. The app then provides the Holder with the DID, leaving the owner of the account ready for testing and certification.

**Verifiers:** Of the three main roles, the process of onboarding Verifiers is the most straightforward. Anyone willing to act as a Verifier can download the app and start verifying. There is no need to create an account for verifying a Holder's certificate. As the Verifier submits no data, the steps of the Verifier onboarding timeline (Fig. 5) do not involve Solid Pods.

*2) Certification:* The certification process requires a Holder to visit an Issuer with the exact document used for identification at the time of onboarding. At this point the Issuer matches this document with the copy stored in the Holder's Solid Pod, viewing it on the app and tapping to accept the ID. The Holder's photo is 'burned' into the upcoming steps so that at the final step of verification, there will be no need to display the same physical ID. In Fig. 6, we see the 'behind the scenes' view of certification, including the Holder's Solid Pod with the ID.

The app is designed to work in a completely decentralized environment. Its functionalities run across the Issuer's, Holder's, and Verifier's phones as well as on the hosting servers, but does not have access to any data from a central database. Every time the app needs to execute an operation, it reads the data from a particular user's Solid Pod (and only with the user's permission). In Fig. 6, at (A) we see that the app reads the allowed identity details from the Holder's Solid Pod, and at (B) compares their hash with the corresponding hash on the blockchain and confirms this on the Issuer's phone display.

Once the identity is confirmed, via physical document checks and Verifiable Credentials demonstrating ownership of the relevant DIDs, the Issuer conducts the antibody test and initiates the process of generating a certificate at (C). A certificate is a set of data in W3C RDF (Resource Description Framework) format [20] containing the test results and a Verifiable Credential for the just-tested Holder. While the hash of the certificate goes onto the blockchain at (D), the original document resides in the Solid Pod (E). It is notable that neither the blockchain nor a third-party centralized server stores the personal data of the Holder.

The Holder has the option of keeping a copy of the certificate in a cloud server of his or her choice. In the event of losing the phone, the Holder can retrieve the data from the cloud and restore the certificate in the regenerated local Solid Pod of the replacement phone. This certificate is visible on the Holder's app in the form of a QR code, giving an easy-to-scan option for Verifiers.

*3) Verification:* The innards of Verification are conceptually similar to what we have just shown for Certification and are thus provided separately in the Supplementary Materials.

### F. Benchmark Testing

To anticipate scalability, we benchmarked three operations (Issuing, Verifying, and Uploading) against a baseline ping that simply echoed a response following a request.

For both Issuing and Verifying we used two variants, to assess the difference between generating hashes (a) locally within the mobile phone app and (b) externally on a server before adding to the blockchain. The Uploading times are purely the times for uploading a certificate to a Solid Pod stored in the cloud, in case that is the Holder's preference.

## III. RESULTS

### A. COVID-19 Antibody Test Certification: App Characteristics

Our 'COVID-19 antibody test certification' (CAT/VC) app builds upon the Verifiable Credentials and Solid frameworks described in Section II, plus our own expertise developed over the past 5 years in the area of blockchain-based certification [21], [22]. The result combines the following characteristics:
- Wholly resident on the end-user's smartphone, yet usable as a paper-only certificate in appropriate socio-economic contexts, as described in the Supplementary Materials.
- One-tap scan, display, and verification of antibody test results, which are owned by the user.
- The app only reveals verifiable CAT/VC results without revealing any personally sensitive information, at the discretion of the user.
- The details of Verifiable Credentials, Solid Pods, and Ethereum blockchain are hidden: from the user's point of view, it is 'just another app'.

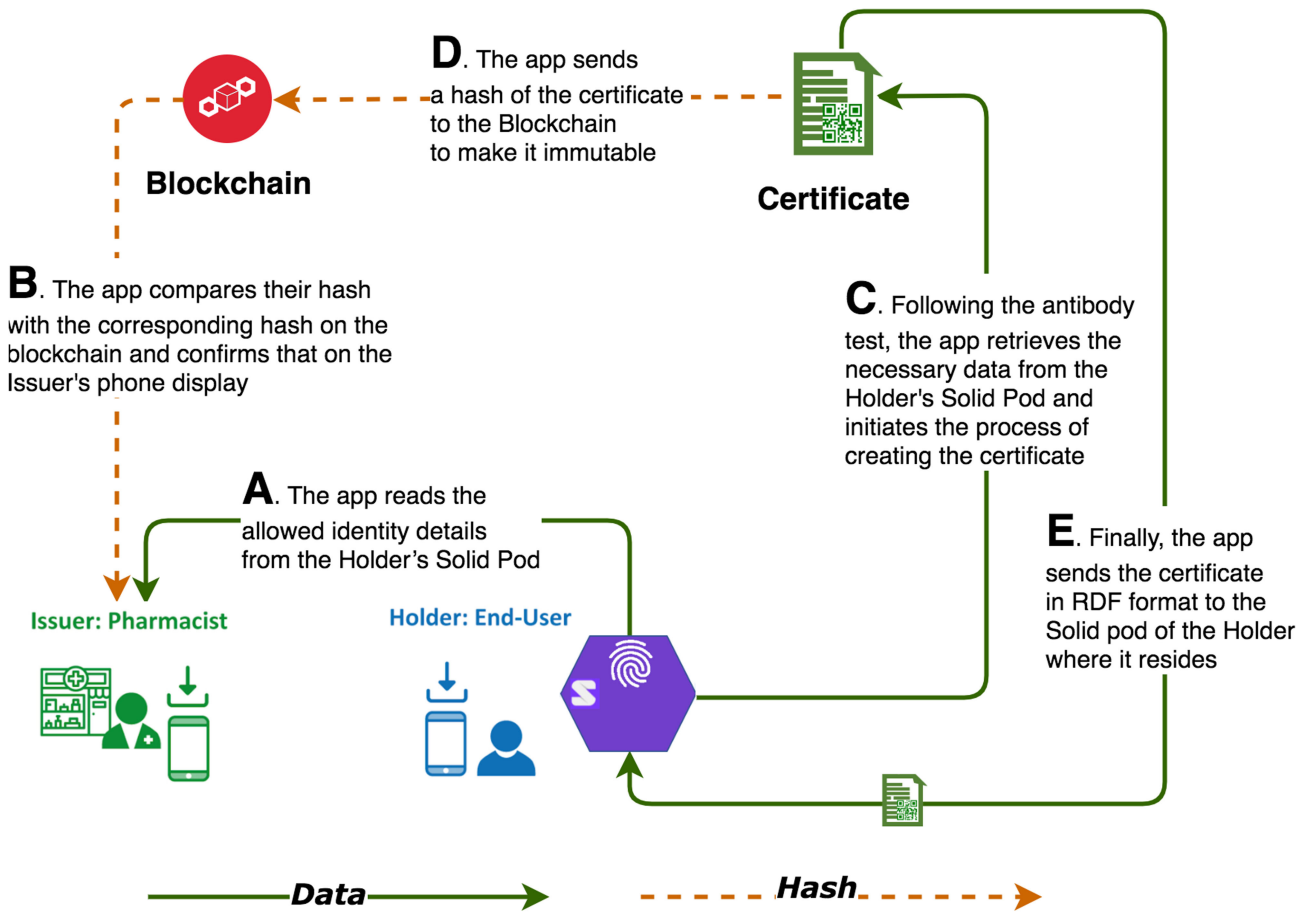

Fig. 6. Certification: main dataflows.

### B. Performance Benchmarking Results

Fig. 7 shows the time to completion in seconds (Y axis) of all six operations where we sent between 1 and 100 simultaneous requests (X axis): the fastest (baseline) ping is the lowest line. Uploading is the second least expensive operation, while Verifying and Issuing are the two most expensive operations of our app. The relative difference in time between operations involving locally generated hash (LH) and server-generated hash (SH) is modest for Issuing (6.9% difference between 'Issuing SH' and 'Issuing LH'), but more twice that for Verifying (17.1% difference between 'Verifying SH and 'Verifying LH'). This behavior is understandable, as Issuing requires writing on the blockchain through transactions (i.e. the method that allows adding an entry to the distributed ledger) while Verifying involves only a look up at a particular ledger entry.

Linear growth for all operations indicates that our architecture is capable of handling scale-up without surprise: there is simply no inter-node or inter-app communication or interaction overhead, so by improving the configuration of the common infrastructures in the architecture, such as any Solid cloud server, blockchain node, or any other intermediate element, the architecture can serve more parallel requests, i.e. reduce the response time. Implications and additional results are discussed below and in the Supplementary Materials.

## IV. DISCUSSION

### A. Deployment, Integration, and Scale

Our focus is on trusted certification, and for this reason we remain committed to deployments involving nationally approved locations such as pharmacies or UK National Health Service surgeries rather than home testing (off-site lab tests are described in the Supplementary Materials). With our approach, deploying the decentralized servers requires another dozen or so Consortium members in addition to the five already engaged, plus about two days of training, which can be handled in parallel for all Consortium members via webinars, as we already do in our current work with blockchain-based educational certification [23]. The mobile phone app itself requires just a download and less than 30 minutes of training for Issuers, and even less for Verifiers and Holders—we anticipate developing a video tutorial for all scenarios. More significantly is the 'buy-in' i.e. acceptance by certified pharmacies and, in the UK, the National Health Service, and integration with existing work practice, ethics guideline approval, and agreement about what, if any, data needs to be stored centrally (no central storage is required at all by our approach). For a full-scale rollout, it would be necessary to further stress-test our prototype along the lines we have already started as described in the preceding Sections.

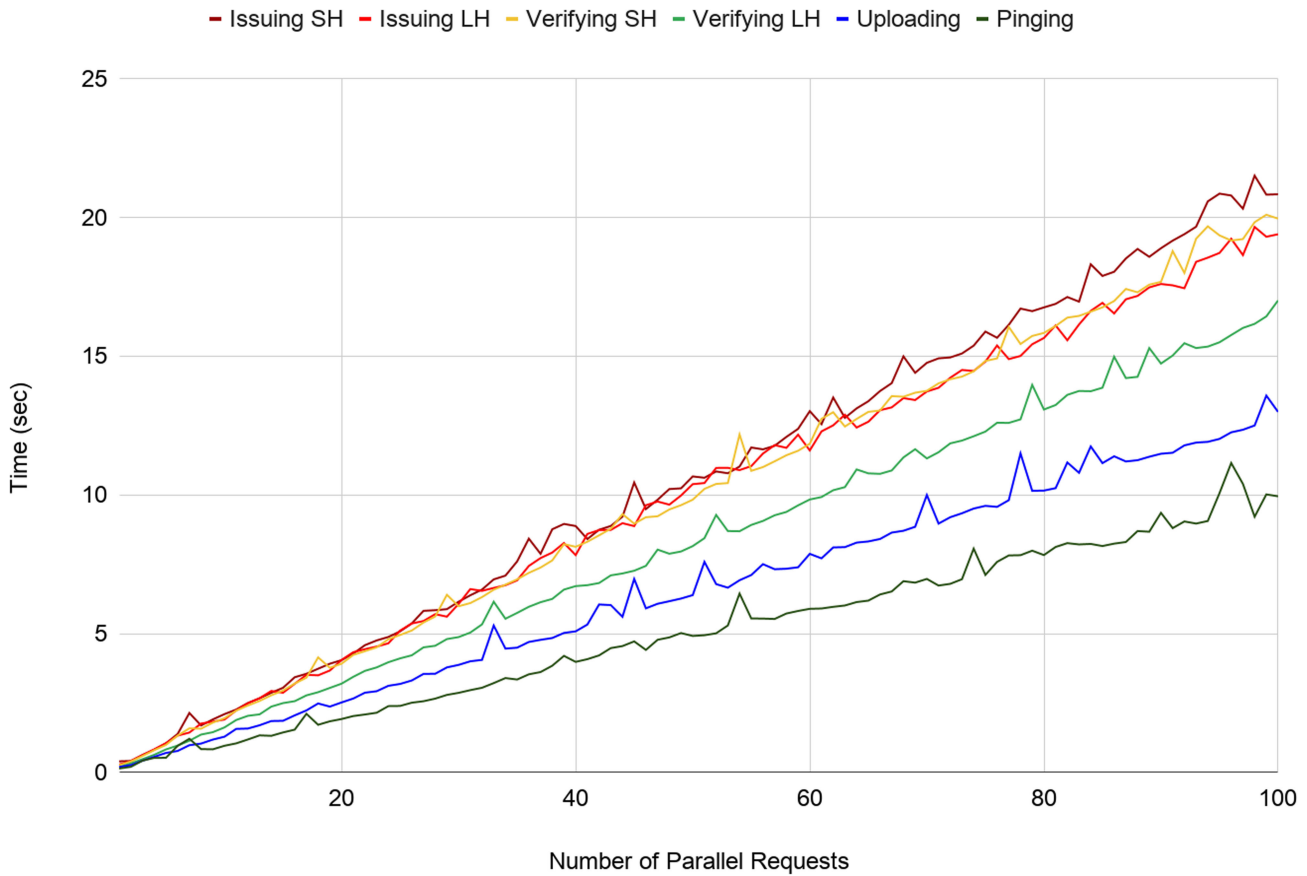

**Fig. 7.** Time to issue up to 100 parallel requests for 'Issuing' (SH=Server Hash and LH=Local Hash), 'Verifying' (SH=Server Hash and LH=Local Hash) and Uploading of Solid Pod data vs baseline standard 'Ping'.

The technology itself is inherently scalable as our Results section shows: transactions on the Consortium blockchain typically take under 5 *seconds* to be confirmed after entry by the Issuer, after which other steps such as verification are subjectively instantaneous. This scales well, as the architecture is inherently distributed across servers (blockchain nodes) and mobile phone apps. Moreover, we have shown worst case results, covering the case when (a) all Solid servers are hosted on the same machine, (b) all blockchain transactions are being sent to the same specific node, and (c) all users are acting simultaneously. In the best case scenario, all simultaneous users would connect to their own Solid servers, and any simultaneous blockchain transactions would each involve different blockchain nodes, so performance overhead would be constant for each additional user. Realistically, i.e. in between these cases, there would be ample numbers of Solid servers, many dozens to hundreds of blockchain nodes, and natural spacing between transactions, and thus performance overhead for each additional user would be minimal.

Collaborative possibilities for rollout and integration are promising, as new initiatives in this niche are rapidly emerging [24].

### B. Beyond Antibody Test Certification

Our scenario highlights antibody testing, but the technology is identical for vaccination certification, as we describe in the Supplementary Materials — this may prove even more popular once vaccines have been suitably tested and approved [6], [7]. The app and decentralized server architecture are readily scalable and applicable generically. For example,

- People could demonstrate that they are eligible to use different methods of transport or to visit public places such as libraries, theaters, or holiday destinations.

- Utility/building/repair staff seeking access to a place of residence, even in 'normal' healthy times, could 'prove their roles'.
- More generally, the entire area of 'Decentralized Verifiable Personal Health Records', as described in [25], particularly if augmented by the W3C Verifiable Credentials standard [12], can benefit from the approach described herein.

### C. Ethics

New technologies bring new challenges for society. Commentators have argued (e.g. in [8], [9], [26]), that certification of the type we have envisaged, even when totally private and tamper-proof, would entail multiple risks, notably: (a) disenfranchising the poor and others who do not have access to the technology or the tests, or have access but 'fail' the test, and (b) becoming a stepping-stone for future governments to deploy the same concept either to enable or to enforce discrimination based on immunity and other arbitrary conditions. To avoid this technology becoming 'weaponized' for discriminatory purposes, we advocate several measures including optional rather than mandatory use, adherence with UK NHS Information Governance guidelines [27], [28] and oversight by an Ethics Committee. This issue is analyzed in detail in the Supplementary Materials.

## V. CONCLUSION

The perceived need for a COVID-19 Antibody Test / Vaccination Certificate, if shown to be biologically robust and to conform to proposed ethical guidelines, has motivated us to develop a mobile phone app based around Verifiable Credentials, distributed storage of cryptographic public/key pairs, and the decentralized verification of data with confidentiality. This has enabled us to provide a facility that is 'just another app' from the viewpoint of the end-user, healthcare professionals, employers and other relevant authorities — thereby providing a tamper-proof record owned entirely by the end-user, and allowing the end-user selectively to reveal solely the proof of test results without surrendering other personal information (e.g. age, address, blood type, other discovered antibodies or immune deficiencies or other inadvertent revelations in the data set, for which certificate Holders may have no idea how this information might be used by someone else in the future), and requiring only mobile phone app downloads from everyone in the loop. This app and its secure digital certificate thus become a powerful adjunct/enhancement to traditional paper-based certification from the NHS or Pharmaceutical testing authorities — and without the need for the costly installation of special 'e-ticket reader' hardware: the same mobile phone app is sufficient for the task at hand, regardless of which of the three roles is involved. Many other uses of secure and private certification via mobile phone app and decentralized servers are additionally made possible, and our infrastructure can be embedded into any other app or web portal through APIs.

### ACKNOWLEDGMENT

The authors would like to thank Ben Hawkridge, Pasquale Iero, Michelle Bachler, Kevin Quick, and Harriett Cornish of KMi, plus Open University Professor of Biology David Male for timely advice about immunology, Dr. Elias Ekonomou of Condatis for raising the prospect of Verifiable Personal Health Records, and external readers Mia Eisenstadt and Zaid Hassan for their input regarding ethical considerations.

# Supplementary Materials

# COVID-19 Antibody Test / Vaccination Certification There's an app for that

Marc Eisenstadt, Manoharan Ramachandran, Niaz Chowdhury, Allan Third, John Domingue*

## I. INTRODUCTION

IN this supplementary materials section we provide the following extras: Introduction — more about the premise of immunity; Methods — (a) how we achieve robust privacy, (b) more details about how Verification works and the functional architecture and mobile phone app infrastructure, (c) scenario variations for (i) Issuing a Digital Certificate Without Photo ID, (ii) Issuing a Paper Certificate, (iii) Off-Site Testing Via an External Lab, and (iv) Vaccination + Certification; Results — additional aspects of system performance; Discussion — further observations about rollout and ethical issues.

**The premise of immunity:** Throughout most of the COVID-19 pandemic, the World Health Organisation (WHO) has advocated a 'test-isolate-trace' approach [1]. In parallel, there has been a worldwide cooperative effort to develop a vaccine [2] and to develop numerous serological tests for the presence of antibodies [3]. If immunity is strongly implied by the outcomes of these latter tests, then individuals could be allowed to get back to work, particularly in healthcare and other key areas [4], [5]. The WHO initially warned that the very premise of COVID-19 immunity was itself uncertain [6]. Yet the fast pace of research is already showing promising signs that early testing was flawed, the presence of antibodies in recovered individuals has been confirmed, and re-infection now seems increasingly unlikely [7], [8]. True, some immunologists have argued that COVID-19 immunity could be very weak, because 'reinfection is an issue with the four seasonal coronaviruses that cause about 10% to 30% of common colds' [9]. But others in that same discussion argue that immunity could be valid for 'a year or two', a view shared by Male, who with Golding and Bootman has written a clear exposition on the life-cycle of infection, antibody detection, and likely immunity to COVID-19 [10]. A related challenge is the *quality* of the testing: test *sensitivity* (% positive detection for the right antibodies, so high sensitivity means few false positives) and *specificity* (% negatives correctly detected, so high specificity means few false negatives) are undergoing great scrutiny even as we write this [11], and are naturally a matter of concern, because they must be sufficiently high to make the approach worthwhile. In the meantime, our research aims to find an approach to achieve highly robust certification, so that it is ready to deploy as-and-when the ongoing biological research satisfies the necessary quality criteria.

## II. METHODS

*A. The design of robust privacy*

Several important guidelines concerning privacy were set out by the Sovrin Foundation, a nonprofit organisation with over 70 corporate partners including IBM, Cisco and others, which has the aim of 'driving greater interoperability and a new trust model for securely sharing private information' [12]. We adopt a variation of the three principles set out in the Sovrin.org White Paper [13], modifying their item 2 as shown below.

*1) Pairwise-unique DIDs and public keys*

As Sovrin.org explains, *'Imagine that when you open a new account with an online merchant, instead of giving them a credit card number or phone number, you gave them a DID created just for them. They could still use this DID to contact you about your order, or to charge you a monthly subscription, but not for anything else. If […] your DID were compromised in any way, you would just cancel it and give them a new one—without affecting any other relationship. [consequently…] a pairwise-pseudonymous DID is not worth stealing.'* [13]

*2) Minimum and Encoded Data Storage / User's Choice*

According to [13], *no* private data should be stored on the ledger, even in hashed form, to make it future-attack-proof. Sovrin accepts, as do we, the need for pseudonymous identifiers (DIDs), pseudonymous public keys, and agent addresses (e.g. the mobile phone app endpoints) to be stored in a decentralized ledger, but in addition we offer the user a *choice* regarding whether and where to host personal information (mobile phone, favorite cloud provider, or both), plus the barest minimum for verification purposes, namely *hashes* (irreversible encodings) of private data. This has the following benefits:

- Serves as a user-storage 'vault' for later recovery in case of loss.
- This 'vault' (i.e. the Solid Pod) can reside on the user's phone, or on a favorite cloud provider, or both — it is always the user's choice.
- To facilitate later independent verification, it uses a blockchain with distributed nodes run by a Consortium of trusted providers so that there is neither a single point of failure nor a single 'owner' even of the hash of the certificate.

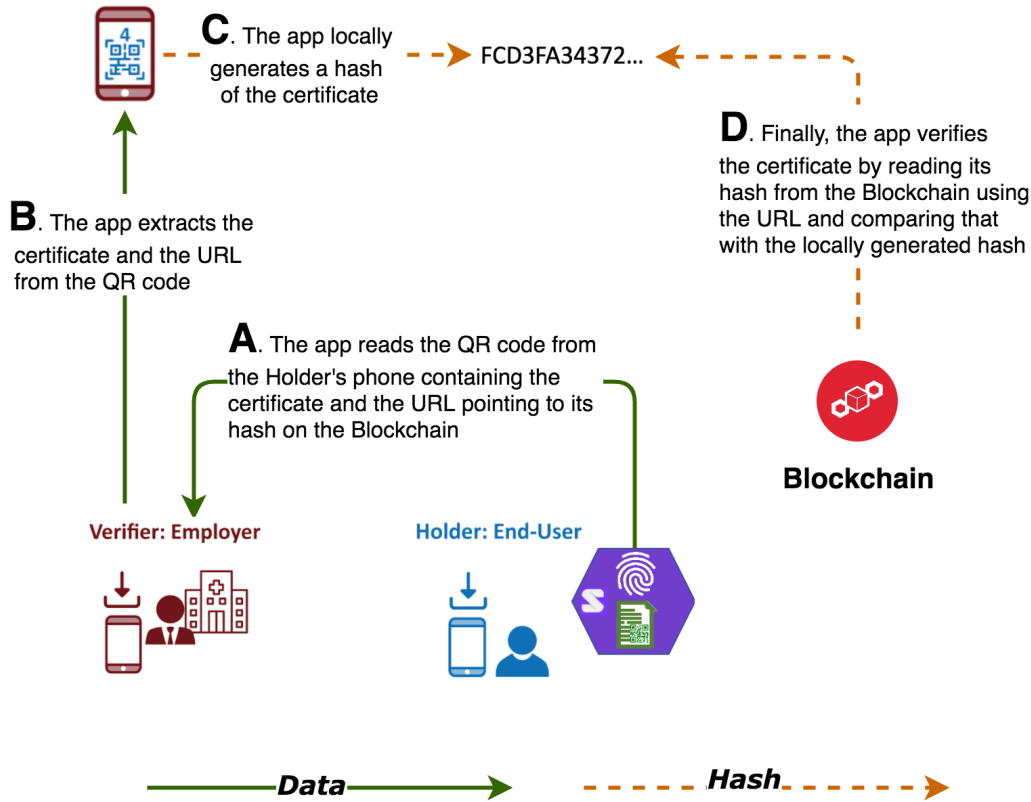

Fig. S1. Verification: main dataflows.

- Even so, it only stores a hash on the Consortium blockchain — a non-reversible but provably correct encoding of the certificate rather than the certificate itself.

This is a powerful privacy-preserving and tamper-proof approach that we call Minimum and Encoded Data Storage / User's Choice. Verborgh [14] has a deeper discussion of the nature and importance of these types of emerging paradigm shifts.

*3) Selective disclosure*

It is essential that users (certificate Holders) should only have to reveal just the portions of their own personally-held private data that are relevant to specific transactions (e.g. proving that you are 18 years of age or older, in order to make certain purchases or access certain locations, but without revealing your actual age or date of birth). This is made possible by the technology known as *cryptographic zero knowledge proofs* [15–17], so named because they provide, to the Verifier who wishes to know, proof of something specific (such as 'Age $\geq$ 18'), but with the Verifier having no knowledge of any other details, in this case actual age or date of birth. The 'secret sauce' of zero knowledge proofs, as illustrated in [16], [17], is that a mathematical function works through a proof of some fact (such as age being greater than or equal to X, or the existence of a certain credential), in such a way that the actual steps involved in executing the proof only reach a positive outcome if the fact is true (for example, the positive outcome may require a certain number of steps to execute): so the proof is valid, but still only indirect (e.g. counting the steps executed) without touching the raw data [15], [16].

*B. Verification and implementation details*

This section describes the operations that underpin the functioning of *verification,* as well as the overall implementation infrastructure and mobile phone app.

*1) Verification*

The process of verifying a certificate is an on-demand action. A Verifier cannot validate a certificate unless requested. It requires a Holder to go to a Verifier for this purpose. A Verifier can be an employer or other individual or organisation to whom the Holder wants or needs to present the certificate. Fig. S1 shows the main data flows involved in Verification.

In Fig. S1, we see that once requested, at (A), the app reads the QR code from the Holder's phone. This QR code (which is generated from the data that itself is stored in the Solid Pod) has two components: the certificate and a URL pointing to the hash on the blockchain. At (B), the app extracts these components and at (C) locally generates a temporary hash of the certificate. Finally (D), the app fetches the hash stored on the blockchain and compares it with the local hash. The matching of the hashes indicates the validity and the authenticity of the certificate stored in the Solid Pod of the Holder. At the same time, the physical identity of the Holder can be confirmed by the Verifier via the Holder's photo ID which will already have been 'burned' into the mobile phone app certificate. The digital identity of the Holder can be confirmed by verifying the Verifiable Credential (embedded in the certificate) based on the relevant Holder DID.

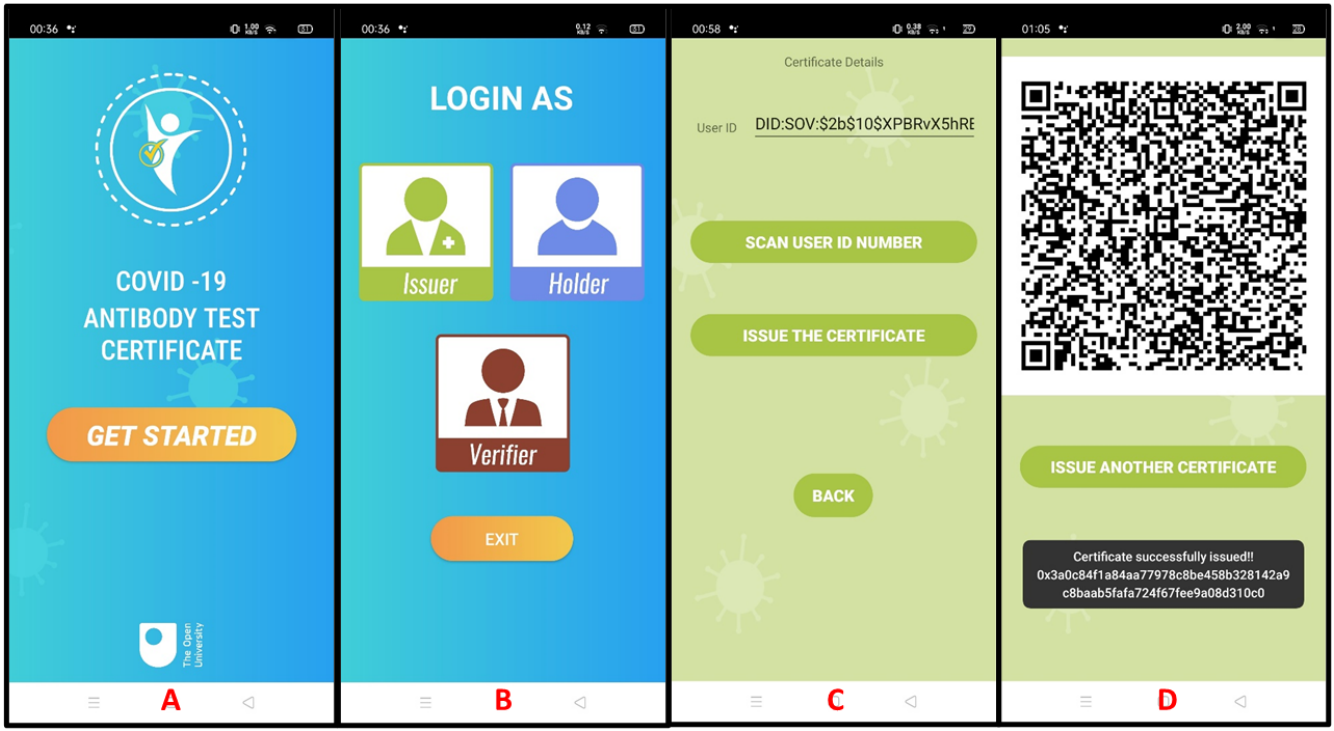

Fig. S2. Representative screen shots of the running mobile app showing (A) home screen, (B) multiple routes for login for the three main roles, just about to tap on 'Issuer', (C) about to issue the certificate having already scanned the user's ID number, displayed at the top, (D) certificate QR code, ready to be scanned by the Holder's mobile phone app.

*2) The functional infrastructure*

The components of our implementation communicate with each other via current Web standards — Hypertext Transfer Protocol Secure (HTTPS), RDF (primarily in the JSON-LD format), Verifiable Credentials, and Decentralized Identifiers — and via blockchain protocols (specifically, Ethereum protocols). The volumes of data and computational requirements are typically small and can be handled by a mobile device (full blockchain nodes are an exception, due to the potential size of the full chain data).

The main software functions required by the implementation are as follows:

*Generate QR codes:* Implemented using standard libraries to generate QR codes for identity and immunity certificates.
*Generate hashes:* Using standard libraries, certificates are transformed into a canonical RDF format before hashing, in order to ensure robust reproducibility of hashes, for verification.
***Communicate with Blockchain:*** The Parity library is used to communicate with our Consortium blockchain. A light client library can handle read/write interactions with the blockchain without requiring a phone to maintain a full copy of the blockchain.
***Communicate with Solid Pods:*** Communication with Solid takes place using the Solid REST API [18], to read and write personal data regarding the Holder to and from their Solid Pod with user permission.
***Manage Issuer and Holder Credentials:*** Issuer and Holder credentials are stored in public/private key wallets containing DIDs. The authorization for an Issuer to create certificates can be represented as a Verifiable Credential issued by the relevant regulatory authority to the Issuer, which any participating party can verify. Currently we use Streetcred ID [19] to generate DIDs for the Issuers, Holders and Certificates.
***Generate Verifiable Credentials:*** Certificates are created at issue time, and their contents asserted as the Claim elements in Verifiable Credentials to be stored in the Holder's Solid Pod, with metadata describing the relevant blockchain records forming the Proof. This provides a sharable data structure which permits anyone to check its authenticity.

*3) The mobile phone app*

Fig. S2 shows representative screen shots of the mobile phone app, which provides all the necessary UI elements for the Issuer, Holder and Verifier to perform their actions. At the time of writing, the main functionalities of the mobile phone app include the ability to scan and generate QR codes and generate hashes for text and images. For the QR code scan and generate functions to work, the mobile phone app is packed with necessary libraries to support QR code functions and only works on smartphones with built-in camera functionality. The mobile phone app also contains the hashing libraries. As the mobile phone app needs to communicate with a server, an active internet connection is necessary for HTTPS server calls.

For speed of implementation for the current prototype, a Node.js Express server does all the heavy lifting for the app, with the functionalities explained above. This is a temporary solution, however, given the urgency of the current situation.

*C. Scenario variations*

Throughout the paper we have focused on a scenario involving 'On-Site Test for Antibodies + Issuance of Digital Certificate Including Photo ID', but there are some key variations easily incorporated into our design, namely (i) 'Issuing Digital Certificate Without Photo ID', (ii) 'Issuing Paper Certificate', (iii) 'Off-Site Testing Via External Lab', and (iv) 'Vaccination + Certification', described in turn below.

*1) Variation 1: Issuing Digital Certificate Without Photo ID*

In our scenario in the main paper, Fig. 2, the Issuer (Pharmacy) needs to authenticate that the Holder is who they say they are, and thus requests that the Holder display both a physical ID, such as a Driving License or a Passport and also a QR code which is scanned by the Issuer using the Issuer's mobile phone app. At this point there is in fact a choice: the Issuer can either (a) tap to accept the ID, in which case the Holder's photo will be 'burned' into the upcoming steps so that at the final step of verification, there will be no need to display the same physical ID, or (b) leave the Holder to display the physical ID once again at verification time.

If path (b) is chosen, there are other implications. At Verification time, to avoid someone else impersonating the Holder, the Holder must present not only the certificate, but also some proof of identity. In this variation, the Verifier can confirm the identity of the Holder by visually inspecting a physical ID card, and separately scanning the Holder's presented QR code (without ID incorporated) to verify just the certificate.

*2) Variation 2: Issuing Paper Certificate*

At step 2 of our main scenario, the test certificate can in fact be provided purely on paper, which has a dual purpose for the Holder: (a) a fallback in case of mobile phone failure; (b) a 'tech-agnostic' option which enables us to provide certification in a more appropriate manner for cases of socio-economic deprivation. This alternative means that some of the advantage of digital certification will be missing, but the use of printed QR codes which include the image of the Holder are still a useful advance over plain paper certificates. It also provides an alternative for individuals with little access to technology, but for whom a paper-based QR code printout can serve as a 'good enough' and 'effectively tamper-proof' certificate.

*3) Variation 3: Off-Site Testing Via External Lab*

It is likely that in many cases, particularly where large volume or high-quality serology testing is required, the Holder's blood sample has to be sent to a separate lab for processing. In this variation, the Pharmacist can issue a certificate that is flagged as being in a 'pending' state. The lab technician will also have a login to the app, via an additional button on the login screen, and see the list of pending certificates waiting for processing and approval. Once the lab technician has the results for a blood sample, the technician has to scan the QR code attached to the sample (this incorporates the Holder's digital ID, but with no personal information exposed to the lab technician) and then tap a button to issue the certified results to the relevant Holder. At this point, the Holder receives a notification with details of the certified result.

Note that the steps in this variation are just like the steps in 'supply chain provenance' gaining increasing traction in the blockchain 'farm-to-fork' world, typified by the IBM Food Trust [20]. Such efforts are also gaining ground in the area of vaccine supply chain provenance [21]. At each step of the chain, each participant adds the information pertinent to their niche, and digitally signs, while cross-checking automatically for authenticity of provenance at earlier steps in the supply chain. For blood samples, both the issuer and lab technician would add serial numbers and details for the blood sample and containers, syringes as necessary, and respective registration numbers / IDs for their roles as pharmacist and lab technician. At Verifier stage, and even for the lab test manufacturer, similar procedures would be deployed so that the integrity of the whole testing life cycle was ensured.

*4) Variation 4: Vaccination + Certification*

Although the most forward-looking variation (because vaccine research, development, approval, and deployment may take the longest [2]), it fits very smoothly into our existing scenario life cycles. Essentially, the Issuer as described throughout the main section of the paper becomes the person administering the vaccination jab (as opposed to taking a blood sample), and certifying that this has happened in the same manner described for the antibody test certificate. The approach to 'supply chain provenance' discussed in the preceding paragraph also applies to this variation, because the Issuer will have to include details of the vaccination source and batch within the certificate.

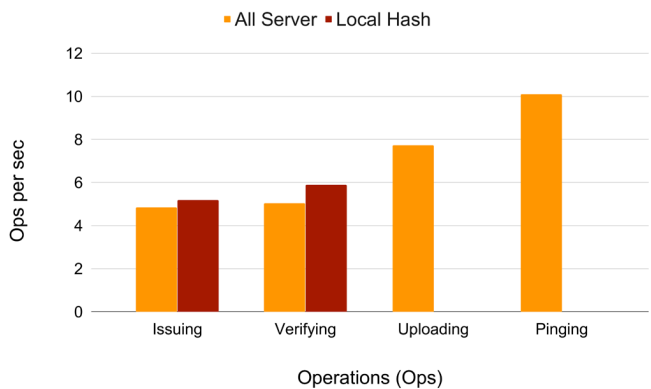

Fig. S3. Operations per second for Issuing (Server vs Local hashing), Verifying' (Server vs Local hashing), Uploading and (baseline) 'Pinging'.

## III. RESULTS

Fig. S3 shows the number of operations per second (Ops/sec) for Issuing, Verifying, Uploading, and Pinging, calculated from the slope of the 1-100 parallel operations timing described in the main paper. It demonstrates that while the current configuration is constant, our architecture can serve about five certificate issuances per second. For verifications, although we experimented with both local and server variants, in practice the hash will be generated locally (within the mobile phone app), giving us the ability to verify about six certificates per second with the existing infrastructure.

This observation shows us that the operations of Issuing and Verifying are twice as expensive as the simplest server ping. Except for some common infrastructure, the architecture is decentralized, i.e. one issuer issues (or verifier verifies) one certificate using one smartphone at a time even if we have hundreds of thousands of parallel requests. Even some commonly held infrastructure can be more distributed, such as the Solid pods. In this experiment, we used just one Solid cloud server for all requests, but in practice, users will have their Solid pod hosted on multiple servers or their own mobile phone. Therefore, if only those common and fixed infrastructures are scaled up, or load-balancing is applied to divert requests over multiple machines, performance time will significantly improve, with a concomitant speedup of Issuing and Verification not requiring architectural re-design.

## IV. DISCUSSION

### A. More about rollout

The architecture presented in the main paper and Supplementary Material above is all built on standard library modules, and therefore joining a Consortium blockchain to help roll this out at scale is relatively straightforward, subject to suitable testing and deployment. The key hurdles are primarily Issuer credentials and the critical mass of the Consortium blockchain. In the case of Issuer credentials, we mentioned in section II.E.1 about Onboarding that we use two factor authentication for Issuers, and an API provided by the General Pharmaceutical Council to cross-check registration — this of course is subject to approval, and relevant discussions are already underway. As for the Consortium blockchain, a strong Consortium of industrial and academic partners needs to be established, after which addition of new members is just a matter of approval by the existing Consortium and the distribution of training and instruction materials. Alternatively, 'parallel' consortia can be created by cloning our approach. Given related ongoing work [22] that we mentioned in the main paper, we are optimistic that critical mass can be achieved.

### B. Ethical considerations

It should be clear from the previous sections that the concepts underlying Verifiable Credentials and the Decentralized Verification of Data with Confidentiality are diametrically opposed to any kind of central data storage or 'Big Brother'-style snooping and data collection, and indeed provide excellent and agreed standards for avoiding such snooping and data collection. To be clear, in the approach we advocate in this paper,

> *Personally identifiable information is stored entirely under the Holder's control (on a mobile phone, on the Holder's cloud provider of choice, or both), and additionally for later verification purposes in minimal (a few bytes) encoded form (hash) on a Consortium blockchain. Moreover, the app allows the user selectively to present only the specific test result, with no other personal information revealed.*

How is it possible that no personal information is stored in a database? What about the certificate itself? That's the beauty of Verifiable Credentials, Zero Knowledge Proofs and our approach of Minimum and Encoded Data Storage / User's Choice: taken together, this combined approach offers cryptographically signed, verifiable, un-tamperable proof that the certificate being shown was really granted by a known testing authority to the person in question, even without showing the name, address, phone number or even UK NHS number of the person holding it.

Everything in this app is decentralized. Anyone wishing to abandon involvement in this kind of certification can just delete the Verifiable Credentials stored on their Solid Pods. There will be no records whatsoever, as if they had never been on the system. Deleting data on the Solid Pods will also turn the hashes on the blockchain into 'orphans' (no data pointing to the hash), i.e. the hashes will become meaningless: it is not possible to recover the original data from a hash.

This almost-too-good-to-be-true approach does raise a fresh concern, raised briefly in the main paper: the same techniques we are advocating seem to open up what we call the *'Private Verifiable Credentials Paradox'*: your digital mobile phone app certificate is so much more private and tamper-proof than the old paper or database versions that it *could* (deliberately or accidentally), be weaponized *for discrimination against your fellow citizens*. In other words, a potential problem, according to critics, is not that the architecture is too weak, but that it is too strong.

Clearly, the more powerful methods of today and tomorrow have the potential to open up a Pandora's Box of Bad Use, if not by the modern democracies in which we may have grown up, then by *some* authority in another time or place - as the world has witnessed all too tragically in the past. We started this project with the noble aim of facilitating a way to get people back to work and heading towards recovery from the devastating impact of the Coronavirus Pandemic of 2019/2020. If COVID-19 antibodies can indeed be shown reliably to confer immunity, and the overwhelming support for the 'test-test-test' mantra of the World Health Organization continues to hold, then people *are* going to get tested, in overwhelming numbers, and certificates *are* going to be issued in one form or another.

But we are not adopting a 'give-up-and-accept-our-fate-in-the-hands-of-bad-actors' approach. Yes, a secure digital certificate could hypothetically be weaponized to a greater degree than a paper one, but the actual degree could be something of a mind-set illusion. *Any* certification method has such potential, and therefore, rather than casting the technology in terms of 'good vs evil' we think our approach is best considered as something that involves a trade-off between (a) the advantages of getting people back to work using good privacy-preserving fraud-prevention methods and (b) the disadvantages of discriminatory (mis)use of such methods. Our approach to this trade-off is strongly to nudge things towards (a), and therefore we propose the following concrete steps to achieve this:

- App usage should be strictly opt-in/optional: a paper certificate must always be allowed by default, just as with, say, train or airline tickets. This helps introduce the concept and technology in a gentle manner: people will ultimately decide what they prefer for themselves.
- Implementations must comply with UK NHS Information Governance (IG) guidelines [23], [24]. Compliance should in principle be straightforward, because (a) in our approach, personally identifiable information is stored entirely under the Holder's control, and additionally for later verification purposes in minimal hash-encoded form on a Consortium blockchain, and (b) the app allows the user selectively to present only the specific test result, with no other personal information revealed. Even so, the UK NHS IG documents provide a strong guiding framework for ensuring continuing compliance, particularly with respect to relevant EU GDPR requirements such as 'Right to erasure' and 'Right to data portability': our architecture by its very design avoids database storage of personally identifiable information, but oversight of possible misuse/abuse of this and related technologies needs to be maintained, as the next three bullet points suggest.
- COVID-19 Antibody Test Certificates should only be applied to workers in healthcare and other comparable key sectors, as defined by the appropriate UK Parliamentary process (for example, the list of key exceptions to mandatory business closure during the current pandemic was specified by the UK Ministry of Housing, Communities, and Local Government), with input from an Ethics Committee mentioned next.
- An Ethics Committee, comparable in scope and composition to the UK NHS Research Ethics Committees, should have oversight of actual deployment of the approach advocated herein.
- The approach should be reviewed on a 3-monthly basis.

In a timely and thoughtful analysis of the ethical complexities surrounding COVID-19 antibody test certificates, Persad and Emanuel [25] argue convincingly for the label 'immunity-based licenses' (rather than 'immunity passports') as a way to focus on the positive benefits granted to those who have been infected with COVID-19, without necessarily worsening the lives of those who have not been infected.

Ethical standards are challenging to uphold, but uphold them we must: we see a strong emphasis on ethics as the best way to negotiate a path towards a 'pandemic end game' in a manner acceptable to the widest possible audience.

V. CONCLUSIONS

Will such an app be suitable as part of a 'pandemic exit strategy' for helping get people back to work in key sectors? There are many issues to be addressed first, including the rigorous scrutiny and approval of antibody tests, likelihood and longevity of immunity, agreement concerning ethical oversight, and acceptance by the public. Our approach is intended to ensure that the procedures for creating tamper-proof, verifiable, privacy-preserving certificates are 'ready to go' while waiting for antibody/immunity tests to achieve the required state of robustness and acceptance. We believe that, just as with train e-tickets, end-users will 'vote with their feet' and deploy the app in large numbers once its benefits have been demonstrated. To take a stance against what we call the 'Pandora's Box of Bad Use', we proposed ethical guidelines at the end of the Discussion, which we believe are essential for the principled development and deployment of the prototype described in this paper.